\documentclass[12pt]{iopart}

\usepackage{iopams}
\usepackage{graphicx}
\begin{document}

\title[Author guidelines for IOP journals in  \LaTeXe]{Properties of $Q\bar{Q}$ $(Q \ \epsilon \ b, c)$ mesons in Coulomb plus Power potential}

\author{Bhavin Patel and P.C.Vinodkumar}

\address{Department of Physics, Sardar Patel University, Vallabh Vidyanagar- 388 120, Gujarat, INDIA
} \ead{azadpatel2003@yahoo.co.in}
\begin{abstract}
The decay rates and spectroscopy of the $Q \bar Q$ $(Q \in c, b)$
mesons are computed in non-relativistic phenomenological quark
antiquark potential of the type $V(r)=-\frac{\alpha_c}{r}+A
r^{\nu}$, (CPP$_{\nu}$) with different choices $\nu$. Numerical
solution of the schrodinger equation has been used to obtain the
spectroscopy of $Q\bar{Q}$  mesons. The spin hyperfine, spin-orbit
and tensor components of the one gluon exchange interaction  are
employed to compute the spectroscopy of the few lower $S$ and
orbital excited states. The numerically obtained radial solutions
are employed to obtain the decay constant, di-gamma and di-leptonic
decay widths. The decay widths are determined with and without
radiative corrections. Present results are compared with other
potential model predictions as well as with the known experimental
values.
\end{abstract}

%\keywords{Suggested keywords} %Use showkeys class option if keyword
\maketitle

  \def\be{\begin{equation}}      \def\ol{\overline}
  \def\ee{\end{equation}}    \def\beq{\begin{eqnarray}}
  \def\dis{\displaystyle}    \def\eeq{\end{eqnarray}}
  \def\btd{\bigtriangledown}     \def\m{\multicolumn}
  \def\dfac{\dis\frac}       \def\ra{\rightarrow}

\section{Introduction}

Heavy flavour hadrons play an important role in several high energy
experiments as well as in the understanding of the theories like
QCD, NRQCD, pNRQCD, vNRQCD and effective field theories. The BES  at
the Beijing Electron Positron Collider (BEPC), E835 at Fermilab, and
CLEO at the Cornell Electron Storage Ring (CESR) are able to collect
the huge data on heavy flavour mesons. Where as B-meson factories,
BaBar at PEP-II and Belle at KEKB are working on the observation of
new and possibly exotics quarkonia states. The CDF and D$\phi$
experiments at Fermilab measuring heavy quarkonia production from
gluon-gluon fusion in $p \bar p $ annihilations at 2 Te$V$. Also
some other experiments like ZEUS and H1 at DESY are studying
charmonia production in photon-gluon fusion. The study related to
the charmonia production and suppression in heavy-ion collisions are
being looked by PHENIX, STAR and NA60. All these experiments are
capable of observing new states, new production mechanisms, new
decays and transitions, and in general to the collection of high
statistics and precision data sample. In the near future, even
larger data samples are expected from the BES-III upgraded
experiment, while the B factories and the Fermilab Tevatron will
continue to supply valuable data for few years. Later on, the LHC
experiments at CERN, Panda at GSI  etc are capable of offering
future opportunities and challenges in this field of heavy flavour
physics \cite{Brambilla2007}.\\
On the theoretical side, heavy quarkonium  provides  testing and the
validity of perturbative QCD,  potential models and lattice QCD
calculations \cite{Seth2005}. The investigation of the properties of
mesons composed of a heavy quark and antiquark ($c \bar c$, $b \bar
c$, $b \bar b$) gives very important insight into heavy quark
dynamics and to the understanding of the constituent quark masses.
The theoretical predictions of the heavy quarkonia $c \bar c$, $b
\bar c$ and $b \bar b$ mesons have rich spectroscopy with many
narrow states of charmonium lying under the threshold of open charm
production \cite{Branes2005,HQP2005} and of botomonium lying under
the threshold of $B-B$ production. Many of these states have not
confirmed or understood by experiments \cite{PDG2006}. However,
there have been renewed interest in the spectroscopy of the heavy
flavoured hadrons due to number of experimental facilities (CLEO,
DELPHI, Belle, BaBar, LHCb etc) which have been continuously
providing and expected to provide more accurate and new informations
about these states at the heavy flavour sector. \\
At the hadronic scale the non-perturbative effects connected with
complicated structure of QCD vacuum necessarily  play an important
role. All this leads to a theoretical uncertainty in the $Q\bar Q$
potential at large and intermediate distances. It is just in this
region of large and intermediate distance that most of the basic
hadron resonances are formed. So the success  of theoretical model
predictions of most of the hadronic properties with experiments can
provide important information about the quark-antiquark
interactions. Such information is of great interest, as it is not
possible to obtain the $Q \bar Q$ potential starting from the basic
principle of the quantum chromodynamics (QCD) at the hadronic scale. \\
Among many theoretical attempts or approaches to explain the hadron
properties based on its quark structure very few were successful in
predicting the hadronic properties starting from its spectroscopy to
decay rates. The nonrelativistic potential models with Buchm\"uller
and Tye \cite{BuchmullerTye1981}, Martin
\cite{Martin1980,Amartin1979,Richadson1979}, Log
\cite{Quiggrosner1977,QuiggRosner1979}, Cornell \cite{Eichten1978}
etc. were successful in predictions of the spectra of the heavy
flavour mesons while the Bethe-Salpeter approach under harmonic
confinement \cite{vijayakumar2004} were successful at low flavour
sector. Though there exist relativistic approaches for the study of
the different hadronic properties
\cite{Altarelli1982,Ebert2003,EbertMod2003}, the non-relativistic
models
have also been equally successful  at the heavy flavour sector.\\
For the theoretical predictions of different decay rates most of the
models require supplementary corrections such as higher order QCD
effects, radiative contributions etc. Even in some cases rescaling
of the model radial wave functions are also being considered.
However the NRQCD formalism provides a systematic approach to study
the decay properties like the di-gamma and the di-lepton decays.
These partial decay widths provide an account of the compactness of
the qurkonium system which is an useful information complementary to
spectroscopy \cite{Rosner2006}. Thus, in this paper we make an
attempt to study the properties like mass spectrum, decay constants
and other decay properties of the $Q\bar{Q}$ systems $(Q \  \epsilon
\ b ,c)$ based on a phenomenological coulomb plus power potential
(CPP$_{\nu}$). Here, we consider different choices of the potential
power index
$\nu$ to study the properties of the mesonic systems upto few excited states. \\
\section{Nonrelativistic Treatment for Heavy Quarks}
There are many theoretical approaches both relativistic and
nonrelativistic to study the heavy quark systems
\cite{QuiggRosner1979,SNGupta1996,SGogfrey1986,Khadkikar1983,hawng1996,JNPandya2001,Vinodkumar1999,
AKRai2002,AKRai2005,AKRai2006,EL-hady1999,Radford2007}. However
their predictions suggests the successes of the nonrelativistic
treatment for the heavy flavour quark-antiquark system
\cite{QuiggRosner1979}. The relativistic invariant theory for
example light-front QCD \cite{Chienhep-ph/0609036} though deals with
different aspects of QCD, under non-relativistic approximations,
reproduces the results comparable to the non-relativistic
quark-potential models \cite{Chienhep-ph/0609036}. In the center of
mass frame of the heavy quark-antiquark system, the momenta of the
quark and antiquark are dominated by their rest mass $m_{Q, \bar
Q}\gg \Lambda_{QCD}\sim |\vec{p}\ |$, which constitutes the basis of
the non-relativistic treatment. For examples NRQCD formalism for the
heavy quarkonia, the
velocity of heavy quark is chosen as the expansion parameter \cite{hycheng1997}.\\
Hence, for the study of heavy-heavy bound state systems such as $c
\bar c$, $b \bar c$ and $b \bar b$, we consider a nonrelativistic
Hamiltonian given by \cite{AKRai2002,AKRai2005,AKRai2006}
\begin{equation}
\label{eq:nlham} H= M + \frac{p^2}{2M_1} +  V(r)\end{equation}
where
 \begin{equation}
\label{eq:mm1} M=m_Q + m_{\bar Q} , \  \ \  \ and \   \  \  \
M_1=\frac{m_Q \ m_{\bar Q}}{m_Q + m_{\bar Q}}\end{equation}
 $m_Q$ and $m_{\bar Q}$ are the mass parameters of quark and antiquark
respectively, p is the relative momentum of each quark and $V(r)$ is
the quark antiquark potential. Though linear plus coulomb potential
is a successful well studied non-relativistic model for heavy
flavour sector, their predictions for decay widths are not
satisfactory owing to the improper value of the radial wave function
at the origin compared to other models \cite{AKRai2005}. Thus, in
the present study we consider a general power potential with color
coulomb term of the form
 \begin{equation}
 V(r)=\frac{-\alpha_c}{r} + A r^\nu
 \end{equation} \label{eq:403}
as the static quark-antiquark interaction potential. This potential
belong to the special choices of the generality of the potentials,
$V(r)=- C r^{\alpha}+ D r^{\beta}+V_0$
\cite{Sameer2004-5,Motyka1998,xtsong1991} with $V_0=0$ $\alpha=-1$,
$\beta=\nu$. For the present study, the power index range of $0.1
<\nu<2.0$ have been explored. Here, for mesons, $\alpha _c = \dis
\frac{4}{3} \alpha_s $, $\alpha _s$ being the strong running
coupling constant, $A$ is the potential parameter similar to the
string strength and $\nu $ is a general power, such that the choice,
$\nu =1$ corresponds to the coulomb plus linear potential. The
different choices of $\nu$ here, correspond to different potential
forms. In general, the potential parameter $A$ can also be different
numerically and dimensionally  for each choices of $\nu$. In the
present study of heavy-heavy flavour mesons, we employ the numerical
approach \cite{Lucha1999} to
generate the Schr\"{o}dinger mass spectra.\\
\begin{table}
\begin{center}
\caption{The model Parameters employed in the present study: (The
potential strength A for different power index, $\nu$  is given in
GeV$^{\nu+1}$)} \vspace{0.1in} \label{tab:1}
\begin{tabular}{lllll}
\hline $\nu$ &A ($c \bar {c}$) & A ($b \bar {c}$)& A ($b \bar
{b}$)\\
\hline
%0.1 &   0.5535  &   0.5645  &   0.5184  \\
%0.3 &   0.4411  &   0.4729  &   0.4829  \\
0.5 &   0.3630  &   0.4085  &   0.4600  \\
0.7 &   0.3034  &   0.3582  &   0.4430  \\
0.8 &   0.2784  &   0.3366  &   0.4358  \\
0.9 &   0.2559  &   0.3169  &   0.4296  \\
1.0 &   0.2355  &   0.2986  &   0.4237  \\
1.1 &   0.2170  &   0.2817  &   0.4180  \\
1.3 &   0.1846  &   0.2513  &   0.4080  \\
1.5 &   0.1573  &   0.2246  &   0.3984  \\

\hline
\end{tabular}
\\ $\alpha_c(c \bar {c})=0.40$, $\alpha_c(b \bar {c})=0.34$,
$\alpha_c(b \bar {b})=0.30$, \\ $m_c=1.24 \ GeV $ and $m_b=4.50 \
GeV$
\end{center}
\end{table}

\begin{figure}[t]

\begin{center}
\includegraphics[height=3.0in,width=2.9in]{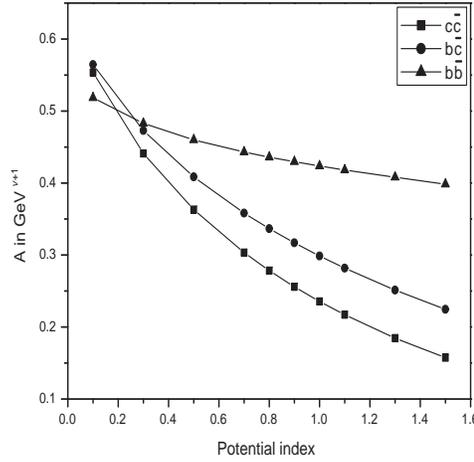}
\vspace{-0.2in} \vspace{-0.2in} \caption{Behavior of $A$ with the
potential index $\nu$ for different $Q \bar Q$
systems}\label{fig:R1s}
\end{center}
\end{figure}
\section{Spin-Dependent forces in $Q \bar Q$ States}
In general, the quark-antiquark bound states are represented by
$n^{2S+1} L_J$,  identified with the $J^{PC}$ values, with $ \vec
J=\vec L + \vec S$, $ \vec S=\vec S_Q + \vec S_{\bar Q}$, parity
$P=(-1)^{L+1}$ and the charge conjugation $C = (-1)^{L+S}$ with (n,
L) being the radial quantum numbers. So the $S$-wave $(L=0)$ bound
states are represented by $J^{PC}=0^{-+}$ and $1^{--}$ respectively.
The $P$-wave $(L=1)$ states are represented by $J^{PC}=1^{+-}$ with
$L=1$ and $S=0$ while $J^{PC}=0^{++},\ 1^{++}$ and $2^{++}$
correspond to $L=1$ and $S=1$ respectively. Accordingly, the
spin-spin interaction among the constituent quarks provides the mass
splitting of $J=0^{-+}$ and $1^{--}$ states, while the spin-orbit
interaction provides the mass splitting of $J^{PC}=0^{++},\ 1^{++}$
and $2^{++}$ states. The $J^{PC} = 1^{+-}$ state with $L=1$ and
$S=0$ represents the center of weight mass of the $P$-state as its
spin-orbit contribution becomes zero, while the two $J=1^{+-}$
singlet and  the $J=1^{++}$ of the triplet P-states form a mixed
state. The $D$-wave $(L=2)$ states are represented by
$J^{PC}=2^{-+}$ with $L=1$ and $S=0$ while $J^{PC}=3^{--},\ 2^{-+}$
and $1^{-+}$ correspond to $L=2$ and $S=1$ respectively. \\
For computing the mass difference between these states, we consider
the spin dependent part of the usual OGEP given by
 \cite{Gerstein1995} as
\begin{equation} V_{S_{\bar {Q}} \cdot S_Q}(r)=\frac{2}{3}
\frac{\alpha_c}{M_{\bar{Q}} m_Q} \ \vec{S_{\bar {Q}}} \cdot
\vec{S_Q} \ 4\pi \delta(\vec{r})  ;  \ \  V_{{L} \cdot S}(r)=
\frac{\alpha_c}{M_{\bar{Q}} m_Q}\ \frac{\vec{L} \cdot \vec{S}}{r^3}
\label{eq:2.4} \end{equation} and
\begin{equation}
  V_{T}(r)= \frac{\alpha_c}{M_{\bar{Q}} m_Q}\
\frac{(3 (\vec S  \cdot{\vec n})(\vec S  \cdot{\vec n}) - \vec{S}
\cdot \vec S)}{r^3},\ \ \ \vec n=\frac{\vec r}{r}
 \label{eq:2.5}  \end{equation}
 The spin average mass for the ground
state is computed for the different choices of $\nu$ in the range,
$0.5 < \nu < 1.5$. The model parameters used here are listed in
Table \ref{tab:1}. The
 potential parameter A are fixed for each choices of $\nu$ so
as to get the experimental ground state spin average masses of $Q
\bar{Q}$ systems. The spin average masses of $c \bar{c}$ is computed
using the experimental ground state mass of $M_{\eta_{c}}=$ 2.980
GeV and $M_{J/ \psi}=$ 3.097 GeV \cite{PDG2006}, while the
experimental values of $M_{\Upsilon}=$ 9.460 GeV and theoretically
predicted values for $\eta_{b}$, $M_{ \eta_{b}}=$ 9.400 GeV
\cite{Ebert2003} are used to get the centre of weight mass of $b
\bar{b}$ system. For the $b \bar{c}$ meson we use the experimental
mass of  $M_{B_{c}}=$ 6.286 GeV \cite{PDG2006} and the theoretically
predicted value of $M_{B_{c}^{*}}=$ 6.332 GeV \cite{Ebert2003}. For
the $nJ$ state, we compute the spin-average or the center of weight
mass from the respective experimental values as
 \begin{equation}
M_{CW,n}=\frac{\sum_{J} 2(2J+1)\ M_{nJ}}{\sum_{J}
2(2J+1)}\end{equation}
\begin{table}
\begin{center}
\caption{The radial Wave function $|R_{ns}(0)|^2$ (in Ge$V^3$) of
$Q\bar{Q}$ systems in various potential models including
CPP$_{\nu}$.}
 \label{tab:2}
\begin{tabular}{llcccccc}
%\multicolumn{10}{}{}\\
\hline
Mesonic&Potential&$1S$&$2S$&$3S$&$4S$&$5S$&$6S$ \\
System&Model&&&&&& \\
\hline
$c\bar{c}$  &   (CPP$_{\nu}),  \nu= $ 0.5  &   0.420   &   0.198   &   0.136   &   0.106   &   0.088   &   0.075   \\
    &   \ \ \ \ \  \ \ \ \ \ \ \ \ \ \ \ \ \ 0.7    &   0.529   &   0.295   &   0.221   &   0.182   &   0.158   &   0.141   \\
    &   \ \ \ \ \  \ \ \ \ \ \ \ \ \ \ \ \ \ 0.8    &   0.577   &   0.347   &   0.270   &   0.229   &   0.202   &   0.183   \\
    &   \ \ \ \ \  \ \ \ \ \ \ \ \ \ \ \ \ \ 0.9    &   0.622   &   0.400   &   0.323   &   0.280   &   0.252   &   0.232   \\
    &   \ \ \ \ \  \ \ \ \ \ \ \ \ \ \ \ \ \ 1.0    &   0.662   &   0.454   &   0.379   &   0.337   &   0.309   &   0.288   \\
    &   \ \ \ \ \  \ \ \ \ \ \ \ \ \ \ \ \ \ 1.1    &   0.700   &   0.509   &   0.439   &   0.399   &   0.372   &   0.352   \\
    &   \ \ \ \ \  \ \ \ \ \ \ \ \ \ \ \ \ \ 1.3    &   0.767   &   0.623   &   0.569   &   0.538   &   0.517   &   0.502   \\
    &   Martin \cite{Martin1980}  &   0.979   &   0.545   &   0.390   &   0.309   &   0.257   &   0.222   \\
    &   Log \cite{Quiggrosner1977}&   0.796   &   0.406   &   0.277   &   0.211   &   0.172   &   0.145   \\
    &   Cornell \cite{Eichten1980} &   1.458   &   0.930   &   0.793   &   0.725   &   0.683   &   0.654   \\
    &   Buchmuller-Tye \cite{BuchmullerTye1981}  &   0.794   &   0.517   &   0.441   &   0.404   &   0.381   &   0.365   \\
    &   Lichtenberg-Wills \cite{Lichenberg1978}  &   1.121   &   0.693   &   0.563   &   0.496   &   0.453   &   0.423   \\
    \hline
$b\bar{c}$  &   (CPP$_{\nu}),  \nu= $ 0.5  &   0.886   &   0.411   &   0.280   &   0.217   &   6.865   &   0.154   \\
    &   \ \ \ \ \  \ \ \ \ \ \ \ \ \ \ \ \ \ 0.7    &   1.109   &   0.609   &   0.454   &   0.373   &   7.221   &   0.288   \\
    &   \ \ \ \ \  \ \ \ \ \ \ \ \ \ \ \ \ \ 0.8    &   1.207   &   0.714   &   0.553   &   0.468   &   7.414   &   0.374   \\
    &   \ \ \ \ \  \ \ \ \ \ \ \ \ \ \ \ \ \ 0.9    &   1.298   &   0.823   &   0.661   &   0.573   &   7.615   &   0.473   \\
    &   \ \ \ \ \  \ \ \ \ \ \ \ \ \ \ \ \ \ 1.0    &   1.381   &   0.933   &   0.776   &   0.688   &   7.823   &   0.587   \\
    &   \ \ \ \ \  \ \ \ \ \ \ \ \ \ \ \ \ \ 1.1    &   1.457   &   1.047   &   0.898   &   0.814   &   8.039   &   0.716   \\
    &   \ \ \ \ \  \ \ \ \ \ \ \ \ \ \ \ \ \ 1.3    &   1.594   &   1.278   &   1.163   &   1.098   &   8.487   &   1.020   \\
    &   Martin \cite{Martin1980} &   1.720   &   0.957   &   0.685   &   0.452   &   0.452   &   0.390   \\
    &   Log \cite{Quiggrosner1977} &   1.508   &   0.770   &   0.524   &   0.401   &   0.325   &   0.275   \\
    &   Cornell \cite{Eichten1980}&   3.191   &   1.769   &   1.449   &   1.297   &   1.205   &   1.141   \\
    &   Buchmuller-Tye \cite{BuchmullerTye1981} &   1.603   &   0.953   &   0.785   &   0.705   &   0.658   &   0.625   \\
    &   Lichtenberg-Wills \cite{Lichenberg1978}  &   2.128   &   1.231   &   0.975   &   0.846   &   0.766   &   0.711   \\
    \hline
$b\bar{b}$  &   (CPP$_{\nu}),  \nu= $ 0.5  &   4.222   &   1.750   &   1.151   &   0.876   &   0.716   &   0.610   \\
    &   \ \ \ \ \  \ \ \ \ \ \ \ \ \ \ \ \ \  0.7    &   5.101   &   2.534   &   1.828   &   1.479   &   1.265   &   1.118   \\
    &   \ \ \ \ \  \ \ \ \ \ \ \ \ \ \ \ \ \  0.8    &   5.487   &   2.948   &   2.214   &   1.840   &   1.607   &   1.443   \\
    &   \ \ \ \ \  \ \ \ \ \ \ \ \ \ \ \ \ \  0.9    &   5.843   &   3.377   &   2.632   &   2.244   &   1.996   &   1.820   \\
    &   \ \ \ \ \  \ \ \ \ \ \ \ \ \ \ \ \ \  1.0    &   6.170   &   3.814   &   3.078   &   2.687   &   2.433   &   2.251   \\
    &   \ \ \ \ \  \ \ \ \ \ \ \ \ \ \ \ \ \ 1.1    &   6.470   &   4.259   &   3.551   &   3.168   &   2.917   &   2.735   \\
    &   \ \ \ \ \  \ \ \ \ \ \ \ \ \ \ \ \ \ 1.3    &   7.006   &   5.173   &   4.575   &   4.249   &   4.034   &   3.877   \\
    &   Martin \cite{Martin1980} &   4.423   &   2.461   &   1.763   &   1.394   &   1.164   &   1.004   \\
    &   Log \cite{Quiggrosner1977}&   4.706   &   2.401   &   1.636   &   1.250   &   1.015   &   0.857   \\
    &   Cornell \cite{Eichten1980}&   14.060  &   5.681   &   1.449   &   3.672   &   3.322   &   3.088   \\
    &   Buchmuller-Tye \cite{BuchmullerTye1981}  &   6.253   &   3.068   &   2.356   &   2.032   &   1.845   &   1.721   \\
    &   Lichtenberg-Wills \cite{Lichenberg1978}  &   6.662   &   3.370   &   2.535   &   2.139   &   1.902   &   1.740   \\

\hline

\end{tabular}
\end{center}
\end{table}

\begin{table}
\caption{Mass spectra (in Ge$V$) of $c\bar{c}$ states.} \label{tab:3}
\begin{tabular}{llllllllllll}
\hline Meson&\multicolumn{6}{c}{\textbf{\underline{\ \ \ \ \ \ \ \ \
\ \
Potential index  $\nu$\ \ \ \ \ \ \ \ \ \ \ \ \ }}}&Expt.&EFG03&ZVR95&BGE05\\
State  &0.7&0.8&0.9&1.0&1.1&1.3&\cite{PDG2006}&\cite{Ebert2003}&\cite{Zeng1995}&\cite{Branes2005} \\

\hline
$1 ^3S_1$   &   3.091   &   3.093   &   3.095   &   3.097   &   3.099   &   3.101   &   3.097   &   3.096   &   3.100   &   3.090   \\
$1^1S_0$    &   2.999   &   2.993   &   2.987   &   2.982   &   2.977   &   2.968   &   2.980   &   2.979   &   3.000   &   2.982   \\
    &       &       &       &       &       &       &       &       &       &       \\
$1 ^3P_2$   &   3.450   &   3.488   &   3.523   &   3.557   &   3.589   &   3.647   &   3.556   &   3.556   &   3.540   &   3.556   \\
$1 ^3P_1$   &   3.419   &   3.451   &   3.481   &   3.508   &   3.535   &   3.582   &   3.511   &   3.510   &   3.500   &   3.505   \\
$1 ^3P_0$   &   3.403   &   3.432   &   3.459   &   3.484   &   3.508   &   3.549   &   3.415   &   3.424   &   3.440   &   3.424   \\
$1 ^1P_1$   &   3.434   &   3.469   &   3.502   &   3.533   &   3.562   &   3.614   &       &   3.526   &   3.510   &   3.516   \\
    &       &       &       &       &       &       &       &       &       &       \\
$2 ^3S_1$   &   3.457   &   3.519   &   3.580   &   3.641   &   3.700   &   3.815   &   3.686   &   3.686   &   3.730   &   3.672   \\
$2^1S_0$    &   3.406   &   3.459   &   3.511   &   3.562   &   3.611   &   3.707   &   3.654   &   3.588   &   3.670   &   3.630   \\
    &       &       &       &       &       &       &       &       &       &       \\
$1 ^3D_3$   &   3.683   &   3.751   &   3.817   &   3.879   &   3.939   &   4.051   &       &   3.815   &   3.830   &   3.806   \\
$1 ^3D_2$   &   3.694   &   3.764   &   3.832   &   3.898   &   3.960   &   4.078   &       &   3.813   &   3.820   &   3.800   \\
$1 ^3D_1$   &   3.677   &   3.744   &   3.808   &   3.869   &   3.927   &   4.036   &   3.770   &   3.798   &   3.800   &   3.785   \\
$1 ^1D_2$   &   3.686   &   3.754   &   3.820   &   3.883   &   3.944   &   4.057   &       &   3.811   &   3.820   &   3.799   \\
    &       &       &       &       &       &       &       &       &       &       \\
$2 ^3P_2$   &   3.662   &   3.755   &   3.847   &   3.939   &   4.030   &   4.209   &       &   3.972   &   4.020   &   3.972   \\
$2 ^3P_1$   &   3.639   &   3.727   &   3.814   &   3.900   &   3.985   &   4.150   &       &   3.929   &   3.990   &   3.925   \\
$2 ^3P_0$   &   3.628   &   3.713   &   3.797   &   3.880   &   3.962   &   4.120   &       &   3.824   &   3.940   &   3.852   \\
$2 ^1P_1$   &   3.651   &   3.741   &   3.831   &   3.919   &   4.007   &   4.179   &       &   3.945   &   3.990   &   3.954   \\
    &       &       &       &       &       &       &       &       &       &       \\
$3 ^3S_1$   &   3.673   &   3.784   &   3.897   &   4.011   &   4.125   &   4.355   &   4.040   &   4.088   &   4.180   &   4.072   \\
$3^1S_0$    &   3.635   &   3.737   &   3.841   &   3.945   &   4.049   &   4.256   &       &   3.991   &   4.130   &   4.063   \\
    &       &       &       &       &       &       &       &       &       &       \\
$4 ^3S_1$   &   3.833   &   3.987   &   4.146   &   4.309   &   4.475   &   4.816   &   4.415   &       &       &   4.406   \\
$4^1S_0$    &   3.801   &   3.947   &   4.097   &   4.250   &   4.406   &   4.723   &       &       &       &   4.384   \\
    &       &       &       &       &       &       &       &       &       &       \\
$5 ^3S_1$   &   3.962   &   4.155   &   4.356   &   4.564   &   4.780   &   5.229   &       &       &       &       \\
$5^1S_0$    &   3.935   &   4.120   &   4.312   &   4.511   &   4.716   &   5.140   &       &       &       &       \\
                                                                                    \\
$6^3S_1$    &   4.073   &   4.300   &   4.540   &   4.792   &   5.055   &   5.609   &       &       &       &       \\
$6^1S_0$    &   4.049   &   4.269   &   4.500   &   4.742   &   4.994   &   5.522   &       &       &       &       \\

\hline
\end{tabular}
\end{table}

\begin{table}
\begin{center}
\caption{Mass spactra (in $GeV$) of $b\bar{c}$ states.}
\label{tab:4}
\begin{tabular}{llllllllllll}
%\multicolumn{10}{}{}\\
\hline Meson&\multicolumn{6}{c}{\textbf{\underline{\ \ \ \ \ \ \ \ \
\ \
Potential index  $\nu$\ \ \ \ \ \ \ \ \ \ \ \ \ }}}&AEH05&EFG03&ZVR95&& \\
 State&0.7&0.8&0.9&1.0&1.1&1.3&\cite{EL-hady2005}&\cite{Ebert2003}&\cite{Zeng1995}& \\
\hline
$1 ^3S_1$   &   6.328   &   6.329   &   6.330   &   6.331   &   6.332   &   6.333   &   6.416   &   6.332   &   6.340   \\
$1^1S_0$    &   6.283   &   6.280   &   6.278   &   6.275   &   6.273   &   6.269   &   6.380   &   6.270   &   6.260   \\
    &       &       &       &       &       &       &       &       &       \\
$1 ^3P_2$   &   6.700   &   6.737   &   6.772   &   6.804   &   6.835   &   6.891   &   6.837   &   6.762   &   6.760   \\
$1 ^3P_1$   &   6.685   &   6.719   &   6.752   &   6.781   &   6.809   &   6.860   &   6.772   &   6.749   &   6.740   \\
$1 ^3P_0$   &   6.678   &   6.711   &   6.741   &   6.770   &   6.797   &   6.845   &   6.693   &   6.734   &   6.730   \\
$1 ^1P_1$   &   6.693   &   6.728   &   6.762   &   6.793   &   6.822   &   6.876   &   6.775   &   6.699   &   6.680   \\
    &       &       &       &       &       &       &       &       &       \\
$2 ^3S_1$   &   6.709   &   6.770   &   6.831   &   6.890   &   6.949   &   7.063   &   6.896   &   6.881   &   6.900   \\
$2^1S_0$    &   6.684   &   6.741   &   6.798   &   6.852   &   6.906   &   7.011   &   6.875   &   6.835   &   6.850   \\
    &       &       &       &       &       &       &       &       &       \\
$1 ^3D_3$   &   6.947   &   7.016   &   7.083   &   7.146   &   7.208   &   7.322   &   7.003   &   7.081   &   7.040   \\
$1 ^3D_2$   &   6.952   &   7.022   &   7.090   &   7.155   &   7.218   &   7.335   &   7.000   &   7.079   &   7.030   \\
$1 ^3D_1$   &   6.944   &   7.013   &   7.079   &   7.142   &   7.202   &   7.315   &   6.959   &   7.077   &   7.020   \\
$1 ^1D_2$   &   6.948   &   7.017   &   7.085   &   7.148   &   7.210   &   7.325   &   7.001   &   7.022   &   7.010   \\
    &       &       &       &       &       &       &       &       &       \\
$2 ^3P_2$   &   6.918   &   7.011   &   7.103   &   7.194   &   7.285   &   7.462   &   7.186   &   7.156   &   7.160   \\
$2 ^3P_1$   &   6.908   &   6.998   &   7.087   &   7.176   &   7.263   &   7.435   &   7.136   &   7.145   &   7.150   \\
$2 ^3P_0$   &   6.902   &   6.991   &   7.080   &   7.166   &   7.252   &   7.421   &   7.081   &   7.126   &   7.140   \\
$2 ^1P_1$   &   6.913   &   7.004   &   7.095   &   7.185   &   7.274   &   7.448   &   7.139   &   7.091   &   7.100   \\
    &       &       &       &       &       &       &       &       &       \\
$3 ^3S_1$   &   6.930   &   7.041   &   7.154   &   7.268   &   7.382   &   7.611   &   7.215   &   7.235   &   7.280   \\
$3^1S_0$    &   6.911   &   7.019   &   7.127   &   7.236   &   7.345   &   7.564   &   7.198   &   7.193   &   7.240   \\
    &       &       &       &       &       &       &       &       &       \\
$4 ^3S_1$   &   7.093   &   7.247   &   7.407   &   7.570   &   7.737   &   8.079   &   7.468   &       &       \\
$4^1S_0$    &   7.077   &   7.228   &   7.384   &   7.542   &   7.704   &   8.035   &   7.452   &       &       \\
    &       &       &       &       &       &       &       &       &       \\
$5 ^3S_1$   &   7.224   &   7.418   &   7.620   &   7.829   &   8.046   &   8.498   &       &       &       \\
$5^1S_0$    &   7.211   &   7.401   &   7.599   &   7.804   &   8.015   &   8.455   &       &       &       \\
                                                                            \\
$6^3S_1$    &   7.337   &   7.565   &   7.807   &   8.060   &   8.324   &   8.883   &       &       &       \\
$6^1S_0$    &   7.325   &   7.550   &   7.788   &   8.036   &   8.295   &   8.841   &       &       &       \\
\hline
\end{tabular}
M(${1 ^1S_0 }$)=6.286 Expt. \cite{PDG2006}.
\end{center}
\end{table}

\begin{table}\caption{Mass spectra (in $GeV$) of $b\bar{b}$
states.} \label{tab:5}

\begin{tabular}{llllllllllll}
%\multicolumn{10}{}{}\\
\hline Meson&\multicolumn{6}{c}{\textbf{\underline{\ \ \ \ \ \ \ \ \
\ \ \ \ \ \ \ \
Potential index  $\nu$\ \ \ \ \ \ \ \ \ \ \ \ \ \ \ }}}&Expt.&EFG03&ZVR95& \\
 State&0.7&0.8&0.9&1.0&1.1&1.3&\cite{PDG2006}&\cite{Ebert2003}&\cite{Zeng1995}& \\
\hline
$1 ^3S_1$   &   9.458   &   9.459   &   9.460   &   9.460   &   9.461   &   9.463   &   9.460   &   9.460   &   9.460   \\
$1^1S_0$    &   9.407   &   9.404   &   9.402   &   9.400   &   9.397   &   9.393   &       &   9.400   &   9.410   \\
    &       &       &       &       &       &       &       &       &       \\
$1 ^3P_2$   &   9.855   &   9.891   &   9.926   &   9.958   &   9.989   &   10.046  &   9.913   &   9.913   &   9.860   \\
$1 ^3P_1$   &   9.841   &   9.876   &   9.908   &   9.938   &   9.967   &   10.019  &   9.893   &   9.892   &   9.870   \\
$1 ^3P_0$   &   9.835   &   9.868   &   9.899   &   9.928   &   9.955   &   10.006  &   9.860   &   9.863   &   9.850   \\
$1 ^1P_1$   &   9.848   &   9.883   &   9.917   &   9.948   &   9.978   &   10.033  &       &   9.901   &   9.880   \\

    &       &       &       &       &       &       &       &       &       \\
$2 ^3S_1$   &   9.861   &   9.920   &   9.979   &   10.037  &   10.094  &   10.205  &   10.023  &   10.023  &   10.020  \\
$2^1S_0$    &   9.836   &   9.891   &   9.946   &   9.999   &   10.052  &   10.154  &       &   9.993   &   10.000  \\
    &       &       &       &       &       &       &       &       &       \\
$1 ^3D_3$   &   10.102  &   10.171  &   10.238  &   10.302  &   10.363  &   10.479  &       &   10.162  &   10.150  \\
$1 ^3D_2$   &   10.106  &   10.176  &   10.244  &   10.309  &   10.371  &   10.489  &   10.162  &   10.158  &   10.150  \\
$1 ^3D_1$   &   10.099  &   10.168  &   10.234  &   10.298  &   10.358  &   10.473  &       &   10.153  &   10.140  \\
$1 ^1D_2$   &   10.103  &   10.172  &   10.239  &   10.303  &   10.365  &   10.481  &       &   10.158  &   10.150  \\
    &       &       &       &       &       &       &       &       &       \\
$2 ^3P_2$   &   10.075  &   10.165  &   10.256  &   10.346  &   10.435  &   10.611  &   10.269  &   10.268  &   10.280  \\
$2 ^3P_1$   &   10.066  &   10.154  &   10.242  &   10.330  &   10.416  &   10.587  &   10.255  &   10.255  &   10.260  \\
$2 ^3P_0$   &   10.061  &   10.148  &   10.236  &   10.322  &   10.407  &   10.575  &   10.232  &   10.234  &   10.240  \\
$2 ^1P_1$   &   10.070  &   10.160  &   10.249  &   10.338  &   10.425  &   10.599  &       &   10.261  &   10.270  \\
    &       &       &       &       &       &       &       &       &       \\
$3 ^3S_1$   &   10.083  &   10.191  &   10.301  &   10.412  &   10.523  &   10.748  &   10.355  &   10.238  &   10.290  \\
$3^1S_0$    &   10.065  &   10.169  &   10.275  &   10.381  &   10.488  &   10.703  &       &   10.355  &   10.370  \\
    &       &       &       &       &       &       &       &       &       \\
$4 ^3S_1$   &   10.244  &   10.394  &   10.549  &   10.709  &   10.871  &   11.207  &   10.279  &       &       \\
$4^1S_0$    &   10.229  &   10.375  &   10.527  &   10.682  &   10.840  &   11.165  &       &       &       \\
    &       &       &       &       &       &       &       &       &       \\
$5 ^3S_1$   &   10.373  &   10.560  &   10.757  &   10.961  &   11.173  &   11.616  &   10.865  &       &       \\
$5^1S_0$    &   10.360  &   10.544  &   10.737  &   10.937  &   11.144  &   11.576  &       &       &       \\
                                                                            \\
$6^3S_1$    &   10.482  &   10.703  &   10.938  &   11.185  &   11.443  &   11.991  &   11.019  &       &       \\
$6^1S_0$    &   10.471  &   10.688  &   10.920  &   11.163  &   11.416  &   11.952  &       &       &       \\

\hline

\end{tabular}
\end{table}
In the case of quarkonia ($c \bar c$ and $b \bar b$ systems)many
orbital excited states are known. Theoretical predictions of all
these states and their decay widths are also being studied. But in
many cases, the decay widths and the spin splitting between
different $J$ values are not well reproduced. Both the decay widths
and the level splitting of the spectra due to the one gluon exchange
interaction terms are related to the values of the radial wave
function or its derivatives at the origin. Thus, the inappropriate
description of the $Q \bar Q$ radial wave function led to the
disparity among the different model predictions of the decay widths
and level splitting. In some cases, for better predictions of the
excited spectra of quarkonia, the strong running coupling constant
$\alpha_{s}$ are evaluated interms of the average kinetic energy of
the quark-antiquark pair at a given state. Accordingly, different
excited states corresponds to have different values of $\alpha_{s}$
\cite{Gerstein1995,Brambilla2002}. However, the radial wave
functions are found to be less sensitive to the changes in
$\alpha_{s}$ compared to similar changes in the values of the
strength of the confining part of the potential. Hence, in this
paper, we allow  $A$ to vary mildly with radial quantum number
$(n=0,1,2...)$ as $A=\frac{A}{(n+1)^{\frac{1}{4}}}$. The variation
in $A$ can be justified by similar arguments for the changes in
$\alpha_{S}$ with the average kinetic energy. Here, as the system
get excited, the average kinetic energy increases and hence the
potential strength (the spring tension) reduces. With this mild
state dependence on the potential parameter $A$, we obtain the
excited spectra as well as the right behavior for the radial wave
functions . The computed values of the radial wave function at the
origin for $(n+1)S$ states are listed in Table \ref{tab:2} for all
the $Q \bar Q$ combinations. Using the spin dependent potential
given by Eqn. \ref{eq:2.4} and \ref{eq:2.5}, we compute the masses
of the different $n^{2 S+1} L_{J}$ states of $c\bar{c}$, $b\bar{c}$
and $b \bar{b}$ mesons. Better stastics with respect to the
experimental values are observed with our predictions of these
states for the potential index lying between 0.7 to 1.3. Thus we
list our predicted properties in this range of potential index only.
The computed masses of the $Q \bar Q$ mesonic states are listed in
Table \ref{tab:3} in the case of $c \bar c$ , in Table \ref{tab:4}
in the case of $b \bar c$ and in Table \ref{tab:5} in the case of $b
\bar b$ systems along with the available experimental values as well
as other model predictions. Fig \ref{fig:R1s} shows the behavior of
$A$ with the potential index $\nu$ that provide us the ground state
center of weight masses for all the three ($c\bar c$, $b\bar c$ and
$b\bar b$) combinations of $Q \bar Q$ systems.
\section{The Decay constants of the heavy flavoured mesons}
The decay constants of mesons are important parameters in the study
of leptonic or non-leptonic weak decay processes. The decay
constants of pseudoscalar ($f_P$) and vector ($f_V$) states are
obtained by prarameterizing the matrix elements of weak current
between the corresponding mesons and the vacuum as \
\begin{equation}
 \langle 0 | \bar Q \gamma^\mu \gamma_5 Q |
P_{\mu}(k)\rangle = i f_P k^\mu
\end{equation}
\begin{equation}
\langle 0 | \bar Q \gamma^\mu Q |
V(k,\epsilon)\rangle = f_V M_V \epsilon^\mu
\end{equation}
where $k$ is the meson momentum, $\epsilon^\mu$ and $M_V$ are the
polarization vector and mass of the vector meson. In the non
relativistic quark model, the decay constant can be expressed
through  the ground state wave function at the origin
$\psi_{P,V}(0)$ by the Van-Royen-Weisskopf formula
\cite{Vanroyenaweissskopf}. The value of the radial wave function
for $0^{- \ +},(R_{P})$ and for $1^{--},(R_{V})$ states would be
different due to their spin dependent hyperfine interaction. The
spin hyperfine interaction of the heavy flavour mesons are small and
this can cause a small shift in the value of the wave function at
the origin. Though, many models neglect this difference between
$(R_{P})$ and $(R_{V})$ we account this correction by considering
 \begin{equation}
R_{nJ}(0)=R(0)\left[1+(SF)_J \frac{<\varepsilon_{SD}>_{nJ}}{M_{1}}
\right]
\end{equation}
Where $(SF)_J$ and $<\varepsilon_{SD}>_{nJ}$ is the spin factor and
spin interaction energy of the meson in the $nJ$ state, while $R(0)$
and $M_{1}$ correspond to the radial wave function at the zero
separation and reduced mass of the $Q \bar Q$ system. It can easily
be seen that this expression is
consistent with the relation \\
\begin{equation}
R(0)=\frac{3 R_{V}+R_{P}}{4}
\end{equation}
given by \cite{Bodwin205}.Though most of the models predict the
mesonic mass spectrum successfully, there are disagreements in the
predictions of their  decay constants. For example, the ratio
$\frac{f_P}{f_V}$ was predicted to be $>1$ as $m_P<m_V$ and their
wave function at the origin $R_P(0) \sim R_V(0)$ by most of the
cases \cite{Hwang1997}. The ratio computed in the relativistic
methods \cite{Wang2006} predicted the ratio $\frac{f_P}{f_V}<1$,
particularly in the heavy flavour sector. The disparity of the
predictions of these decay constants play decisive role in the
precision measurements of the weak decay parameters as well as the
spectroscopic hyperfine splitting. So, we reexamine the predictions
of the decay constants under different potential (by the choices of
different $\nu$) schemes employed in the present work. Incorporating
a first order QCD correction factor, we compute,
\begin{equation}
f^2_{P/V}=\frac{3 \left| R_{nS}(0)\right|^2} { \pi M_{P/V}} {\bar
C^2}(\alpha_s) \label{eq:fpv}
 \end{equation}
here, ${\bar C^2}(\alpha_s)$ is the QCD correction factor given by
\cite{EBraaten1995}
\begin{equation}
{\bar C^2}(\alpha_s)=1+\frac{\alpha_s}{\pi} \left[\frac{m_Q-m_{
q}}{m_Q+m_{q}} \ ln \frac{m_Q}{m_{q}}- \delta^{V,P} \right]
\label{eq:fpvc}
\end{equation}
Where $\delta^{V} = \frac{8}{3} $ and $\delta^P=2$. In the case of
$c \bar c$ and $b \bar b$ systems, ${\bar C^2}(\alpha_s)$ becomes
$1- \frac{\alpha_{s}}{\pi} \delta^{V,P} $  as the first term within
the square bracket vanishes. Our computed values of $f_{P}$ and
$f_{V}$ without this correction and with the correction shown in
brackets up to $6S$ states are tabulated in Tables \ref{tab:6}
-\ref{tab:8} along with available experimental results and with
other theoretical predictions in the cases of $c \bar c$, $b \bar c$
and $b \bar b$ systems respectively.
\begin{table*}
\begin{center}
\caption{Pseudoscalar meson  decay constant $f_P$ (Me$V$), Vector
meson decay constant $f_V$ (Me$V$) and  $f_P/f_V$ of $c\bar{c}$
states (The bracketed quantities are with QCD corrections).}
\label{tab:6}
\begin{tabular}{llllllll}
\hline
\m{0}{c}{}&\m{0}{c}{CPP$_{\nu}$}&\m{0}{c}{1S}&\m{0}{c}{2S}&\m{0}{c}{3S}&\m{0}{c}{4S}&\m{0}{c}{5S}&\m{0}{c}{6S}\\
\hline
$f_{P}$  &   0.7 &   365(295)   &   270(218) &     230(186)   &   206(167)   &   189(153)   &   177(143)   \\
    &   0.8 &   377(305)   &   287(232) &     248(200)   &   224(181)   &   207(168)   &   195(157)   \\
    &   0.9 &   388(314)   &   302(244) &      264(214)   &   241(195)   &   224(181)   &   211(171)   \\
    &   1.0 &   397(321)   &   316(255) &     279(226)   &   256(207)   &   239(193)   &   226(183)   \\
    &   1.1 &   404(327)   &   328(265) &     292(236)   &   269(218)   &   253(205)   &   240(194)   \\
    &   1.3 &   417(337)   &   348(282) &     315(255)   &   293(237)   &   276(224)   &   264(213)   \\
&\cite{Edwards2001}&335$\pm$75\\
&\cite{Cvetic2004}&292$\pm$25\\
\hline
$f_{V}$ &   0.7 &   419(313)   &   291(217)   &   243(181)   &   216(161)   &   197(147)   &   184(137)   \\
    &   0.8 &   439(327)   &   314(234)   &   266(198)   &   238(177)   &   219(163)   &   204(152)   \\
    &   0.9 &   457(341)   &   336(250)   &   287(214)   &   259(193)   &   239(178)   &   225(167)   \\
    &   1.0 &   473(352)   &   356(265)   &   308(230)   &   280(209)   &   260(194)   &   245(182)   \\
    &   1.1 &   487(363)   &   375(280)   &   329(245)   &   300(224)   &   280(208)   &   264(197)   \\
    &   1.3 &   512(382)   &   412(307)   &   367(274)   &   339(253)   &   318(237)   &   303(226)   \\
&\cite{PDG2006}&416$\pm$6&304$\pm$4&187$\pm$8&161$\pm$10\\
&\cite{Cvetic2004}&459$\pm$28\\
&\cite{Wang2006} & 459$\pm$28&364$\pm$24&319$\pm$22&288$\pm$18&265$\pm$16\\
&\cite{EbertMod2003}&551&401&\\
\hline
$\frac{f_{P}}{f_{V}}$ &   0.7 &   0.87(0.94)   &   0.93(1.00)   &   0.95(1.03)   &   0.95(1.04)   &   0.96(1.04)   &   0.96(1.04)   \\
    &   0.8 &   0.86(0.93)   &   0.91(0.99)   &   0.93(1.01)   &   0.94(1.02)   &   0.95(1.03)   &   0.96(1.03)   \\
    &   0.9 &   0.85(0.92)   &   0.90(0.98)   &   0.92(1.00)   &   0.93(1.01)   &   0.94(1.02)   &   0.94(1.02)   \\
    &   1.0 &   0.84(0.91)   &   0.89(0.96)   &   0.91(0.98)   &   0.91(0.99)   &   0.92(0.99)   &   0.92(1.01)   \\
    &   1.1 &   0.83(0.90)   &   0.87(0.95)   &   0.89(0.96)   &   0.90(0.97)   &   0.90(0.99)   &   0.91(0.98)   \\
    &   1.3 &   0.81(0.88)   &   0.84(0.92)   &   0.86(0.93)   &   0.86 (0.94)   &   0.87(0.95)   &   0.87(0.94)   \\
&\cite{PDG2006}&0.81$\pm$0.19\\

\hline

\end{tabular}

 \cite{PDG2006} $\rightarrow$ PDG-2006,  \cite{Edwards2001}$\rightarrow$   Edwards-2001,
  \cite{Cvetic2004}$\rightarrow$ Cvetic-2004,\\ \cite{Wang2006} $\rightarrow$ Wang-2006,  \cite{EbertMod2003}$\rightarrow$Ebert-2003.  \\
\end{center}
\end{table*}
\begin{table*}
\begin{center}
\caption{Pseudoscalar meson  decay constant $f_P$ (Me$V$), Vector
meson decay constant $f_V$ (Me$V$) and  $f_P/f_V$ of $b\bar{c}$
states (The bracketed quantities are with QCD corrections).}
\label{tab:7}

\begin{tabular}{llllllll}
\hline
\m{0}{c}{}&\m{0}{c}{CPP$_{\nu}$}&\m{0}{c}{1S}&\m{0}{c}{2S}&\m{0}{c}{3S}&\m{0}{c}{4S}&\m{0}{c}{5S}&\m{0}{c}{6S}\\
\hline
$f_{P}$ &   0.7 &   396(355)   &   289(260)   &   247(221)   &   222(199)   &   205(184)   &   192(172)   \\
    &   0.8 &   412(370)   &   311(279)   &   270(242)   &   245(220)   &   228(204)   &   215(193)   \\
    &   0.9 &   426(382)   &   331(297)   &   291(261)   &   267(240)   &   250(225)   &   237(213)   \\
    &   1.0 &   439(393)   &   350(314)   &   312(280)   &   289(259)   &   272(244)   &   259(233)   \\
    &   1.1 &   450(403)   &   368(330)   &   332(298)   &   310(278)   &   293(263)   &   281(252)   \\
    &   1.3 &   468(420)   &   401(359)   &   369(331)   &   349(313)   &   334(299)   &   321(288)   \\
    &\cite{Ebert2003}&433\\
    &\cite{Gerstein1995}&460$\pm$60\\
    &\cite{Eichten1994}&500\\

    \hline
$f_{V}$ &   0.7 &   414 (349)   &   296(250)   &   251(212)   &   225(190)   &   207(175)   &   194(164)   \\
    &   0.8 &   432 (364)   &   320(270)   &   276(232)   &   249(210)   &   231(195)   &   218(184)   \\
    &   0.9 &   448 (378)   &   342(288)   &   299(252)   &   273(230)   &   255(215)   &   242(204)   \\
    &   1.0 &   463 (390)   &   363(306)   &   322(271)   &   297(250)   &   279(235)   &   265(224)   \\
    &   1.1 &   476 (401)   &   383(323)   &   344(290)   &   320(270)   &   302(255)   &   289(243)   \\
    &   1.3 &   498 (420)   &   421(355)   &   387(326)   &   364(307)   &   348(293)   &   335(282)   \\
    &\cite{Ebert2003}&503\\
    &\cite{Gerstein1995}&460$\pm$60\\
    &\cite{Eichten1994}&500\\
    \hline
$\frac{f_{P}}{f_{V}}$ &   0.7 &   0.96(1.02)   &   0.98(1.04)   &   0.98(1.04)   &   0.99(1.05)   &   0.99(1.05)   &   0.99(1.05)        \\
    &   0.8 &   0.95(1.02)   &   0.97(1.03)   &   0.98(1.04)   &   0.98(1.05)   &   0.99(1.05)   &   0.99(1.05)        \\
    &   0.9 &   0.95(1.01)   &   0.97(1.03)   &   0.97(1.04)   &   0.98(1.04)   &   0.98(1.05)   &   0.98(1.04)        \\
    &   1.0 &   0.95(1.01)   &   0.96(1.03)   &   0.97(1.03)   &   0.97(1.04)   &   0.97(1.04)   &   0.98(1.04)        \\
    &   1.1 &   0.95(1.00)   &   0.96(1.02)   &   0.97(1.03)   &   0.97(1.03)   &   0.97(1.03)   &   0.97(1.04)        \\
    &   1.3 &   0.94(1.00)   &   0.95(1.01)   &   0.95(1.02)   &   0.96(1.02)   &   0.96(1.02)   &   0.96(1.02)        \\
&\cite{Ebert2003}&0.86\\
 &\cite{Gerstein1995}&1.00\\
 &\cite{Eichten1994}&1.00\\

\hline
\end{tabular}
\cite{Ebert2003}$\rightarrow$ Ebert-2003, \cite{Gerstein1995}
$\rightarrow$ Gerstein-1995, \cite{Eichten1994}$\rightarrow$
Eichten-1994.

 \end{center}
\end{table*}

\begin{table*}
\begin{center}
\caption{Pseudoscalar meson  decay constant $f_P$ (Me$V$), Vector
meson decay constant $f_V$ (Me$V$) and  $f_P/f_V$ of $b\bar{b}$
states (The bracketed quantities are with QCD corrections).}
\label{tab:8}

\begin{tabular}{llllllll}
\hline
\m{0}{c}{}&\m{0}{c}{CPP$_{\nu}$}&\m{0}{c}{1S}&\m{0}{c}{2S}&\m{0}{c}{3S}&\m{0}{c}{4S}&\m{0}{c}{5S}&\m{0}{c}{6S}\\
\hline
$f_{P}$ &   0.7 &   708(606)   &   492(421)   &   414(355)   &   370(317)   &   340(291)   &   318(273)   \\
    &   0.8 &   733(628)   &   528(453)   &   453(388)   &   409(350)   &   379(325)   &   357(306)   \\
    &   0.9 &   756(647)   &   563(482)   &   490(420)   &   448(384)   &   419(359)   &   397(340)   \\
    &   1.0 &   776(665)   &   596(511)   &   527(451)   &   486(416)   &   457(392)   &   436(373)   \\
    &   1.1 &   794(680)   &   627(537)   &   562(481)   &   523(448)   &   495(424)   &   474(406)   \\
    &   1.3 &   824(706)   &   686(587)   &   629(539)   &   594(509)   &   569(488)   &   549(471)   \\
    \hline
$f_{V}$ &   0.7 &   722(584)   &   497(402)    &    417(337)   &   372(301)   &   342(276)   &   320(259)   \\
    &   0.8 &   749(606)   &   534(432)   &   457(369)   &   412(333)   &   382(309)   &   359(291)   \\
    &   0.9 &   773(625)   &   571(462)   &   495(401)   &   452(365)   &   422(341)   &   399(323)   \\
    &   1.0 &   795(643)   &   605(489)   &   533(431)   &   491(397)   &   462(373)   &   439(356)   \\
    &   1.1 &   814(658)   &   638(516)   &   570(461)   &   529(428)   &   501(405)   &   479(388)   \\
    &   1.3 &   847(685)   &   700(566)   &   641(518)   &   605(489)   &   578(468)   &   558(451)   \\
&\cite{PDG2006}&715$\pm$5&498$\pm$5&430$\pm$4&336$\pm$18&369$\pm$42&240$\pm$28\\
&\cite{EbertMod2003}&839&562&\\
&\cite{Wang2006} & 498$\pm$20&366$\pm$27&304$\pm$27&259$\pm$22&228$\pm$16\\
\hline
$\frac{f_{P}}{f_{V}}$ &   0.7 &   0.98(1.04)   &   0.99(1.05)   &   0.99(1.05)   &   0.99(1.05)   &   0.99(1.05)   &   0.99(1.05)   \\
    &   0.8 &   0.98(1.04)   &   0.99(1.05)   &   0.99(1.05)   &   0.99(1.05)   &   0.99(1.05)   &   0.99(1.05)   \\
    &   0.9 &   0.98(1.04)   &   0.99(1.04)   &   0.99(1.05)   &   0.99(1.05)   &   0.99(1.05)   &   0.99(1.05)   \\
    &   1.0 &   0.98(1.03)   &   0.99(1.04)   &   0.99(1.05)   &   0.99(1.05)   &   0.99(1.05)   &   0.99(1.05)   \\
    &   1.1 &   0.98(1.03)   &   0.98(1.04)   &   0.99(1.04)   &   0.99(1.05)   &   0.99(1.05)   &   0.99(1.05)   \\
    &   1.3 &   0.97(1.03)   &   0.98(1.04)   &   0.98(1.04)   &   0.98(1.04)   &   0.98(1.04)   &   0.98(1.04)   \\
\hline

\end{tabular}
 \cite{PDG2006} $\rightarrow$ PDG-2006, \cite{Wang2006}$\rightarrow$Wang-2006, \cite{EbertMod2003}$\rightarrow$ Ebert-2003. \\
 \end{center}
\end{table*}
\begin{table}
\begin{center}
\caption{Mean Square radii ($fm$) for the $Q \bar{Q}$ ($Q
 \  \epsilon$  b,  c)  states in various potential power index, $\nu$.}
 \label{tab:9}

\begin{tabular}{llccccccccc}
%\multicolumn{10}{}{}\\
\hline
&CPP$_{\nu}$&\m{0}{c}{1S}&\m{0}{c}{1P}&\m{0}{c}{2S}&\m{0}{c}{1D}&\m{0}{c}{2P}&\m{0}{c}{3S}&\m{0}{c}{4S}&\m{0}{c}{5S}&\m{0}{c}{6S} \\
&\m{0}{c}{$\nu$}&&&&&&&& \\
\hline
$c\bar{c}$  &   0.7 &   0.50    &   0.79    &   1.07    &   1.03    &   1.33    &   1.59    &   2.08    &   2.55    &   3.00    \\
    &   0.8 &   0.48    &   0.74    &   0.99    &   0.96    &   1.23    &   1.46    &   1.89    &   2.30    &   2.69    \\
    &   0.9 &   0.46    &   0.70    &   0.94    &   0.90    &   1.14    &   1.35    &   1.74    &   2.10    &   2.44    \\
    &   1.0 &   0.45    &   0.67    &   0.89    &   0.85    &   1.07    &   1.26    &   1.61    &   1.93    &   2.23    \\
    &   1.1 &   0.43    &   0.64    &   0.84    &   0.81    &   1.01    &   1.19    &   1.50    &   1.79    &   2.06    \\
    &   1.3 &   0.41    &   0.59    &   0.77    &   0.74    &   0.92    &   1.07    &   1.33    &   1.56    &   1.79    \\
  &\cite{Vinodkumar1999}&0.39&&0.82&&&1.44&2.36&\\
  &\cite{Gunar1997}&0.43&&0.85&&&1.18&1.47&\\
  \hline

$b\bar{c}$  &   0.7 &   0.39    &   0.62    &   0.84    &   0.82    &   1.06    &   1.26    &   1.65    &   2.02    &   2.38    \\
    &   0.8 &   0.38    &   0.58    &   0.79    &   0.76    &   0.97    &   1.16    &   1.50    &   1.83    &   2.14    \\
    &   0.9 &   0.36    &   0.55    &   0.74    &   0.71    &   0.91    &   1.07    &   1.38    &   1.67    &   1.94    \\
    &   1.0 &   0.35    &   0.53    &   0.70    &   0.67    &   0.85    &   1.00    &   1.28    &   1.53    &   1.78    \\
    &   1.1 &   0.34    &   0.51    &   0.67    &   0.64    &   0.80    &   0.94    &   1.19    &   1.42    &   1.64    \\
    &   1.3 &   0.33    &   0.47    &   0.61    &   0.59    &   0.73    &   0.85    &   1.05    &   1.24    &   1.42    \\

\hline
$b\bar{b}$  &   0.7 &   0.25    &   0.41    &   0.55    &   0.54    &   0.70    &   0.83    &   1.10    &   1.35    &   1.60    \\
    &   0.8 &   0.24    &   0.38    &   0.52    &   0.50    &   0.65    &   0.77    &   1.00    &   1.22    &   1.43    \\
    &   0.9 &   0.23    &   0.36    &   0.49    &   0.47    &   0.60    &   0.71    &   0.92    &   1.11    &   1.30    \\
    &   1.0 &   0.22    &   0.35    &   0.46    &   0.45    &   0.57    &   0.66    &   0.85    &   1.02    &   1.19    \\
    &   1.1 &   0.22    &   0.33    &   0.44    &   0.42    &   0.53    &   0.62    &   0.79    &   0.95    &   1.09    \\
    &   1.3 &   0.21    &   0.31    &   0.40    &   0.39    &   0.48    &   0.56    &   0.70    &   0.83    &   0.94    \\

    &\cite{Vinodkumar1999}&0.19&&0.40&&&0.71&1.17&1.85\\
    &\cite{Juan-Luis2008}&0.23&&0.51&&&0.71&0.88&\\
    &\cite{Eichten1980}&0.20&&0.48&&&0.72&0.92&\\
    &\cite{Gunar1997}&0.24&&0.51&&&0.73&0.93&\\
\hline

\end{tabular}

\cite{Vinodkumar1999} $\rightarrow$ Vinodkumar-1999,
\cite{Gunar1997}$\rightarrow$ Gunar-1997 ,\cite{Juan-Luis2008}
$\rightarrow$ Juan-Luis-2008,\\ \cite{Eichten1980} $\rightarrow$
Eichten-1980.\\
\end{center}
\end{table}
\begin{table*}
\begin{center}
\caption{Average quark Velocity in $Q \bar{Q}$ ($Q
 \  \epsilon$  b,  c) states with various potential power index.}
 \label{tab:10}

\begin{tabular}{llccccccccc}
%\multicolumn{10}{}{}\\
\hline
&\m{0}{c}{CPP$_{\nu}$}&\m{0}{c}{1S}&\m{0}{c}{1P}&\m{0}{c}{2S}&\m{0}{c}{1D}&\m{0}{c}{2P}&\m{0}{c}{3S}&\m{0}{c}{4S}&\m{0}{c}{5S}&\m{0}{c}{6S} \\
&\m{0}{c}{$\nu$}&&&&&&&& \\
\hline
$c\bar{c}$:$\left<v_{c}^{2}\right>^\frac{1}{2}$  &   0.7 &   0.245   &   0.263   &   0.268   &   0.297   &   0.289   &   0.295   &   0.320   &   0.341   &   0.360   \\
    &   0.8 &   0.265   &   0.296   &   0.309   &   0.343   &   0.339   &   0.351   &   0.388   &   0.421   &   0.450   \\
    &   0.9 &   0.283   &   0.329   &   0.350   &   0.389   &   0.391   &   0.410   &   0.462   &   0.509   &   0.550   \\
    &   1.0 &   0.300   &   0.361   &   0.392   &   0.435   &   0.445   &   0.472   &   0.541   &   0.603   &   0.660   \\
    &   1.1 &   0.316   &   0.392   &   0.434   &   0.480   &   0.499   &   0.536   &   0.625   &   0.705   &   0.779   \\
    &   1.3 &   0.345   &   0.451   &   0.518   &   0.569   &   0.610   &   0.669   &   0.804   &   0.928   &   1.043   \\
 &\cite{Gunar1997}&0.27&0.29&0.35&0.34&0.39&0.44&0.52&\\
\hline
$b\bar{c}$:$\left<v_{c}^{2}\right>^\frac{1}{2}$ &   0.7 &   0.396   &   0.419   &   0.427   &   0.472   &   0.459   &   0.469   &   0.507   &   0.540   &   0.570   \\
    &   0.8 &   0.427   &   0.472   &   0.492   &   0.544   &   0.539   &   0.558   &   0.616   &   0.667   &   0.713   \\
    &   0.9 &   0.457   &   0.524   &   0.558   &   0.617   &   0.622   &   0.652   &   0.733   &   0.806   &   0.871   \\
    &   1.0 &   0.484   &   0.575   &   0.624   &   0.690   &   0.706   &   0.749   &   0.859   &   0.956   &   1.045   \\
    &   1.1 &   0.509   &   0.624   &   0.691   &   0.762   &   0.793   &   0.851   &   0.991   &   1.118   &   1.234   \\
    &   1.3 &   0.555   &   0.719   &   0.825   &   0.904   &   0.970   &   1.063   &   1.276   &   1.471   &   1.653   \\
 \hline
$b\bar{c}$:$\left<v_{b}^{2}\right>^\frac{1}{2}$  &   0.7 &   0.030   &   0.032   &   0.032   &   0.036   &   0.035   &   0.036   &   0.038   &   0.041   &   0.043   \\
   &   0.8 &   0.032   &   0.036   &   0.037   &   0.041   &   0.041   &   0.042   &   0.047   &   0.051   &   0.054   \\
    &   0.9 &   0.035   &   0.040   &   0.042   &   0.047   &   0.047   &   0.049   &   0.056   &   0.061   &   0.066   \\
    &   1.0 &   0.037   &   0.044   &   0.047   &   0.052   &   0.054   &   0.057   &   0.065   &   0.073   &   0.079   \\
    &   1.1 &   0.039   &   0.047   &   0.052   &   0.058   &   0.060   &   0.065   &   0.075   &   0.085   &   0.094   \\
    &   1.3 &   0.042   &   0.055   &   0.063   &   0.069   &   0.074   &   0.081   &   0.097   &   0.112   &   0.126   \\
 \hline

$b\bar{b}$:$\left<v_{b}^{2}\right>^\frac{1}{2}$ &   0.7 &   0.077   &   0.074   &   0.076   &   0.081   &   0.079   &   0.081   &   0.086   &   0.092   &   0.096   \\
    &   0.8 &   0.083   &   0.083   &   0.087   &   0.094   &   0.093   &   0.096   &   0.105   &   0.113   &   0.121   \\
    &   0.9 &   0.088   &   0.093   &   0.098   &   0.106   &   0.107   &   0.113   &   0.125   &   0.137   &   0.148   \\
    &   1.0 &   0.093   &   0.101   &   0.110   &   0.119   &   0.122   &   0.130   &   0.147   &   0.163   &   0.178   \\
    &   1.1 &   0.097   &   0.110   &   0.122   &   0.132   &   0.137   &   0.147   &   0.170   &   0.191   &   0.210   \\
    &   1.3 &   0.105   &   0.127   &   0.145   &   0.157   &   0.168   &   0.184   &   0.220   &   0.252   &   0.283   \\

 &\cite{Juan-Luis2008}&0.094&&0.091&&&0.103&0.120&\\
 &\cite{Gunar1997}&0.080&0.068&0.081&0.075&0.085&0.096&0.112&\\

\hline

\end{tabular}
 \cite{Juan-Luis2008}
$\rightarrow$Juan-Luis-2008, \cite{Gunar1997}$\rightarrow$ Gunar-1997.\\
\end{center}
\end{table*}
\section{Mean Square Radii and Average quark Velocity of $Q\bar{Q}$ $(Q \epsilon b, c)$ mesons}
Apart from the decay constants, $f_{P/V}$, other important
properties associated with a mesonic state are the mean square radii
$\langle r^2\rangle$ and the mean square velocity of the
quark$/$antiquark $\left<v^{2}_{q}\right>$. The mean square size of
the mesonic states is an important in the estimations of hadronic
transition widths \cite{Gottfried1978,Voloshin1979,Kuang1990} of
different $Q \bar Q^{'}$ systems. The average velocity of the quark
and the antiquark within a $Q \bar Q$ bound state are important for
the estimation of relativistic corrections and are useful
particularly in the NRQCD formalism as well as in the estimation of
the quarkonium production rates \cite{Bodwin2008}. We compute the
mean square radii as
 \begin{equation}
\left\langle r^2\right\rangle= \int_{0}^{\infty} r^4 |R_{nl}(r)|^2
dr \label{eq:meansqradii}
\end{equation}
and the average mean square quark$/$antiquark velocity for the $c
\bar c$ and $b \bar b$ systems, according to the relation given by
\cite{Juan-Luis2008}
\begin{equation}
\left \langle({{v}_{q}}\right)^2\rangle=\frac{1}{2 M_{1}} (E-\langle
V(r)\rangle)\label{eq:velocity}
\end{equation}
Here, $E$ is the binding energy of the system, $M_{1}$ is the
reduced mass of the mesonic system and $\langle V(r)\rangle$ is the
expectation value of the potential. In the $b \bar c$ case, the
velocity of $b$ and $c$ quarks are obtained as
\begin{equation}
\left \langle({{v}_{b}}\right)^2\rangle= (E-\langle V(r)\rangle)
\frac{2\ m_{c}}{m_{b}\ (m_{b}+m_{c})}\end{equation}
\begin{equation} \label{eq:velocity_b}
\left \langle({{v}_{c}}\right)^2\rangle= (E-\langle V(r)\rangle)
\frac{2\ m_{b}}{m_{c}\ (m_{b}+m_{c})} \label{eq:velocity_c}
\end{equation}
 The computed $rms$ radii up to $6S$ states of $c \bar c$, $b \bar
c$ and $b \bar b$ systems are listed in Table \ref{tab:9} for the
range of potential index $0.7\leq \nu \leq1.3$. The estimated $rms$
velocity $\left<v^{2}_{q}\right>^\frac{1}{2}$ of the charm and
beauty quark/antiquark using Eqn. \ref{eq:velocity} to
\ref{eq:velocity_c} are given in Table \ref{tab:10} of $c \bar c$ ,
$b \bar c$ and $b \bar b$ systems in their $1S$, $1P$, $1D$ and $2S$
to $6S$ states.
\begin{table*}
\begin{center}
\caption{ $0^{-+}\rightarrow \gamma \ \gamma$ decay rates (in ke$V$)
of heavy qurkonia states (Bracketed quantities are with QCD
corrections). } \label{tab:11}

\begin{tabular}{llllllll}
%\multicolumn{10}{}{}\\
\hline
\m{0}{c}{}&\m{0}{c}{CPP$_{\nu}$}&\m{0}{c}{1S}&\m{0}{c}{2S}&\m{0}{c}{3S}&\m{0}{c}{4S}&\m{0}{c}{5S}&\m{0}{c}{6S}\\
&\m{0}
{c}{$\nu$}&\m{0}{c}{$\Gamma_{0}$($\Gamma_{\gamma\gamma}$)}&\m{0}{c}{$\Gamma_{0}$($\Gamma_{\gamma\gamma}$)}&
\m{0}{c}{$\Gamma_{0}$($\Gamma_{\gamma\gamma}$)}&\m{0}{c}{$\Gamma_{0}$($\Gamma_{\gamma\gamma}$)}&
\m{0}{c}{$\Gamma_{0}$($\Gamma_{\gamma\gamma}$)}
&\m{0}{c}{$\Gamma_{0}$($\Gamma_{\gamma\gamma}$)}\\
\hline
$c\bar{c}$  &   0.7 &   5.87(3.98)   &   2.83(1.92)   &   1.92(1.30)   &   1.47(1.00)   &   1.20(0.82)   &   1.02(0.69)   \\
    &   0.8 &   6.29(4.26)   &   3.15(2.13)   &   2.17(1.47)   &   1.68(1.14)   &   1.38(0.93)   &   1.17(0.80)   \\
    &   0.9 &   6.65(4.51)   &   3.44(2.33)   &   2.40(1.63)   &   1.87(1.27)   &   1.54(1.04)   &   1.31(0.89)   \\
    &   1.0 &   6.98(4.73)   &   3.70(2.51)   &   2.61(1.77)   &   2.03(1.38)   &   1.68(1.14)   &   1.43(0.97)   \\
    &   1.1 &   7.26(4.92)   &   3.93(2.67)   &   2.79(1.89)   &   2.18(1.48)   &   1.80(1.22)   &   1.53(1.04)   \\
    &   1.3 &   7.74(5.24)   &   4.33(2.93)   &   3.08(2.08)   &   2.40(1.62)   &   1.97(1.33)   &   1.66(1.13)   \\
&\cite{PDG2006}&7.2$\pm$0.7&1.3 $\pm$ 0.6*&\\
&\cite{Lansberg2008}&$7.5-10$&$3.5-4.5$\\
&\cite{Kim2005}&7.14$\pm$0.95&4.44$\pm$0.48\\
&\cite{EbertMod2003}&5.5&1.8\\
\hline
$b\bar{b}$  &   0.7 &   0.44(0.33)   &   0.20(0.15)   &   0.14(0.11)   &   0.11(0.08)   &   0.09(0.07)   &   0.08(0.06)   \\
    &   0.8 &   0.47(0.36)   &   0.23(0.18)   &   0.17(0.13)   &   0.13(0.10)   &   0.11(0.09)   &   0.10(0.07)   \\
    &   0.9 &   0.50(0.38)   &   0.26(0.20)   &   0.19(0.15)   &   0.16(0.12)   &   0.13(0.10)   &   0.12(0.09)   \\
    &   1.0 &   0.53(0.40)   &   0.29(0.22)   &   0.22(0.17)   &   0.18(0.14)   &   0.16(0.12)   &   0.14(0.11)   \\
    &   1.1 &   0.55(0.42)   &   0.32(0.25)   &   0.25(0.19)   &   0.21(0.16)   &   0.18(0.14)   &   0.16(0.12)   \\
    &   1.3 &   0.60(0.45)   &   0.38(0.29)   &   0.31(0.23)   &   0.26(0.20)   &   0.23(0.18)   &   0.21(0.16)   \\
&\cite{Lansberg2008}&$0.56$&$0.269$&0.208\\
&\cite{Kim2005}&0.384$\pm$0.047&0.191$\pm$0.025&\\
&\cite{EbertMod2003}&0.35&0.15&0.1\\
\hline

\end{tabular}

\cite{PDG2006}$\rightarrow$ PDG-2006 ,\cite{Lansberg2008}
 $\rightarrow$Lansberg-2008,  \cite{Kim2005} $\rightarrow$ Kim-2005,\\
\cite{EbertMod2003} $\rightarrow$Ebert-2003 ,* $\rightarrow$Anser-2004 \cite{Anser2004}. \\
\end{center}
\end{table*}

\begin{table*}
\begin{center}
\caption{$1^{--}\rightarrow l^{+}\ l^{-}$ decay rates (in ke$V$) of
heavy qurkonia states (Bracketed quantities are with QCD
corrections).} \label{tab:12}

\begin{tabular}{llllllll}
\hline
\m{0}{c}{}&\m{0}{c}{CPP$_{\nu}$}&\m{0}{c}{1S}&\m{1}{c}{2S}&\m{0}{c}{3S}&\m{0}{c}{4S}&\m{0}{c}{5S}&\m{0}{c}{6S}\\
&\m{0}
{c}{$\nu$}&$\Gamma_{VW}$($\Gamma_{ll}$)&$\Gamma_{VW}$($\Gamma_{ll}$)&$\Gamma_{VW}$($\Gamma_{ll}$)&
$\Gamma_{VW}$($\Gamma_{ll}$)&$\Gamma_{VW}$($\Gamma_{ll}$)&$\Gamma_{VW}$($\Gamma_{ll}$)\\
\hline
$c\bar{c}$  &0.7 &   5.64(2.77)   &   2.44(1.20)   &   1.60(0.79)   &   1.21(0.59)   &   0.97(0.48)   &   0.82(0.40)   \\
    &   0.8 &   6.19(3.04)   &   2.78(1.36)   &   1.85(0.91)   &   1.41(0.69)   &   1.14(0.56)   &   0.96(0.47)   \\
    &   0.9 &   6.69(3.28)   &   3.12(1.53)   &   2.10(1.03)   &   1.61(0.79)   &   1.30(0.64)   &   1.10(0.54)   \\
    &   1.0 &   7.16(3.51)   &   3.45(1.69)   &   2.35(1.15)   &   1.80(0.88)   &   1.47(0.72)   &   1.24(0.61)   \\
    &   1.1 &   7.60(3.73)   &   3.78(1.85)   &   2.60(1.27)   &   1.99(0.98)   &   1.62(0.80)   &   1.37(0.67)   \\
    &   1.3 &   8.38(4.11)   &   4.41(2.17)   &   3.07(1.51)   &   2.37(1.16)   &   1.92(0.94)   &   1.62(0.79)   \\
&\cite{PDG2006}&5.55$\pm$0.14&2.48$\pm$0.06&0.86$\pm$0.07&0.58$\pm$0.07&\\
&\cite{EbertMod2003}&6.7(5.4)&3.2(2.4)&\\
\hline
$b\bar{b}$  &   0.7 &   1.37(0.84)   &   0.62(0.38)   &   0.43(0.26)   &   0.33(0.21)   &   0.28(0.17)   &   0.24(0.15)   \\
    &   0.8 &   1.47(0.91)   &   0.71(0.44)   &   0.51(0.31)   &   0.41(0.25)   &   0.34(0.21)   &   0.30(0.18)   \\
    &   0.9 &   1.57(0.97)   &   0.81(0.50)   &   0.59(0.37)   &   0.48(0.30)   &   0.41(0.25)   &   0.36(0.22)   \\
    &   1.0 &   1.65(1.02)   &   0.90(0.56)   &   0.68(0.42)   &   0.56(0.34)   &   0.48(0.30)   &   0.43(0.26)   \\
    &   1.1 &   1.74(1.07)   &   1.00(0.62)   &   0.77(0.47)   &   0.64(0.39)   &   0.56(0.34)   &   0.50(0.31)   \\
    &   1.3 &   1.88(1.16)   &   1.19(0.74)   &   0.95(0.59)   &   0.81(0.50)   &   0.71(0.44)   &   0.64(0.40)   \\

&\cite{PDG2006}&1.34$\pm$0.018&0.612$\pm$0.011&0.443$\pm$0.008&0.272$\pm$0.029&0.31$\pm$0.071&0.13$\pm$0.03\\
&\cite{EbertMod2003}&1.4(1.3)&0.6(0.5)&\\
\hline
\end{tabular}
\cite{PDG2006} $\rightarrow$ PDG2006, \cite{EbertMod2003}
$\rightarrow$ Ebert-2003.

\end{center}
\end{table*}
\section{Decay rates of quarkonia}
The spectroscopic parameters including the predicted masses and the
resultant radial wave functions are being used here to compute the
decay rates. We consider the conventional Van Royen-Weisskopf
formula for the di-gamma and di-leptonic decay widths. Like in many
other theoretical models, we also consider the contribution from the
radiative corrections to these decays. Accordingly, the two photon
decay width of the pseudoscalar meson is computed as
\begin{equation} \label{eqn:digamma}
\Gamma _{{0^{-+}{\rightarrow}{\gamma \gamma}}}= \Gamma _0 + \Gamma_R
\end{equation}  where $\Gamma_0$ is the conventional Van Royen-Weisskopf formula given by
\cite{Vanroyenaweissskopf}
\begin{equation}\label{eq:0gamma}
 \Gamma_0 = \frac{12 \alpha_e^2 e_Q^4}{M_P^2} \ |R_{nS}(0)|^2
\end{equation}
and $\Gamma_R$ is the radiative correction given by \cite{AKRai2005}
 \begin{equation}
 \Gamma_R=\ \frac{\alpha_s}{\pi} \left(\frac{\pi^2-20}{3}\right) \ \Gamma_0
 \end{equation}
Similarly, the leptonic decay widths of the vector mesons with
radiative correction is computed as
 \begin{equation} \label{eqn:dilepton}
\Gamma _{{1^{--}{\rightarrow}{l^+ l^-}}}=  \Gamma _{VW}
+\Gamma_{rad}
 \end{equation}
where $\Gamma _{VW}$ is the conventional Van Royen-Weisskopf formula
given by
 \begin{equation}
\Gamma _{VW}= \frac {4 \alpha_e^2 e_Q^2}{M_V^2} \ |R_{nS}(0)|^2
\end{equation} \label{eq:vwgamma}
and the radiative correction $\Gamma _{rad}$ is given by
\begin{equation}\Gamma_{rad}=\ -\frac{16}{3 \pi} \alpha_s \ \Gamma
_{VW},\end{equation} It is obvious to note that the computations of
the decay rates and the radiative correction terms described here
require the right description of the meson state through its radial
wave function at the origin R(0) and its mass $M$ which in turn
depend on the model parameters like $\alpha_s$, confinement strength
and quark model masses. Generally, due to lack of exact solutions
for colour dynamics and with the uncertainties over the exact nature
of interquark potential, R(0) and $M$ are also been considered as
free parameters of the theory \cite{Hafsakhan1996}. However, we
found it appropriate to employ the spectroscopic parameters of the
mesons such as the phenomenologically predicted meson mass and the
corresponding wave function predicted by different models for the
estimation of the decay properties of the mesons.\\
Making use of the model parameters, the resultant radial wave
functions and the mesonic mass we compute the $0^{-+}$$\rightarrow $
$\gamma \ \gamma $  and $1^{--}$ $\rightarrow$ $l^+ \ l^- $ decay
widths for each cases of the potential model employed here for the
present study. The results are shown in Table (\ref{tab:11}) for
$0^{-+}$ $\rightarrow $ $\gamma \ \gamma $, and in Table
(\ref{tab:12}) for $1^{--}$ $\rightarrow$ $l^+ \ l^- $ in comparison
with the predictions of the contemporary potential models and with
the known experimental values. The bracketed quantities listed in
both the tables are the decay widths with the respective radiative
corrections added to the conventional V-W formula as per
Eqn.\ref{eqn:digamma} and
 Eqn.  \ref{eqn:dilepton} respectively. \\
\section{Results and Discussion}
We have employed the coulomb plus power potential form to study the
mass spectrum  and decay properties of heavy mesons. Unlike in our
earlier studies using variational approach
\cite{AKRai2002,AKRai2005}, here we solved the Schr\"{o}dinger
equation numerically using \cite{Lucha1999}. It helps us to study
the mass spectrum of $c\bar{c}$, $b\bar{c}$ and $b\bar{b}$ mesons up
to few excited states. Our potential parameters are fixed with
respect to the centre of weight ground state $1S$ mass of the $Q
\bar Q$ $(Q \ \epsilon \ b, c)$ systems. Our predication of the
excited state of these mesons for the potential index $\nu=$ 0.9 to
1.3 are found to be in good agreement with the experimental results
as well as with theoretical predictions of other models.  Success of
the present study is not only related to the numerical approach but
also to the fact that the strength of the confinement part of the
potential is
made state dependent according to the relation $\frac{A}{(n+1)^{\frac{1}{4}}}$. \\
In Table \ref{tab:2}, we tabulate  the values of the $S$-wave radial
wave function at the origin, $|R_{ns}(0)|^2$ (in Ge$V^3$) for the S-
wave of heavy $Q\bar{Q}$ systems along with other models. These
quantities are not only essential inputs for evaluating decay
constants, decay rates, NRQCD parameters and production cross
sections for quarkonium states but also important for the
determination of hyperfine and fine splitting of their mass spectra.
We compared our prediction for the $|R(0)|^2$ with that of Martin
potential \cite{Martin1980}, Logarithmic potential
\cite{Quiggrosner1977}, Cornell potential \cite{Eichten1980},
Buchmuller-Tye potential \cite{BuchmullerTye1981} and
Lichtenberg-Wills potential \cite{Lichenberg1978}. We also observe
that a model independent relationship for the radial wave function
of the $b\bar{c}$ with that of $c\bar{c}$  and $b\bar{b}$ system as
given by \cite{Sterett1997}
\begin{equation}
\left|\psi_{b\bar c}\right|^2\approx\left|\psi_{c \bar
c}\right|^{2(1-q)}\left|\psi_{b\bar b} \right|^{2q}\end{equation}
with $q=0.35$, seem to hold within $2\%$ variation for the lower
states in the potential range $0.5\leq\nu\leq1.5$ and for higher
states we find the relation hold within $5\%$ for all values
for $\nu$ studied here.\\
Our results for the decay constant of pseudoscalar meson $f_{P}$,
vector meson $f_{V}$ and their ratio of $f_{P}/f_{V}$ with and
without the QCD corrections (given in brackets) for $c\bar{c}$,
$b\bar{c}$ and $b\bar{b}$ mesons are listed in Tables \ref{tab:6} to
\ref{tab:8} respectively from $1S$ to $6S$ states. Our results are
compared with the available experimental values \cite{PDG2006} and
with other theoretical predications. We could see that reduction in
the $f_{P}$ values to about $19\%$   in the cases of $c \bar c $,
$14\%$ in the case of $b \bar b$ and $10\%$ in the case of $b \bar
c$ and reduction in the $f_{V}$ values to about $25\%$ in the cases
of $c \bar c$, $19 \%$ in the case of $b \bar b$ and $16\%$
 in the case of $b \bar c$ are attributed due to the QCD
correction factor. Our results for $1S$ state of $f_{P}$ for $c \bar
c$ system is in good agreement with the values reported by CLEO
collaboration  and $f_{V}$ with the PDG average value
\cite{PDG2006}. The ratio $f_{P}/f_{V}$  without the QCD correction
predicted by us lie between 0.87 to 0.8 in the potential range of
$0.7\leq\nu\leq1.3$ as against the experimental ratio of
0.81$\pm$0.19 \cite{PDG2006}. The predicted values of $f_{P}$ for
$2S$ to $6S$ states are in accordance with other theoretical
predictions. Our results for the $c\bar{c}$ meson decay constants
without the QCD corrections are in good agreement with the
experimental data, while that for $b \bar b$ system with the QCD
corrections are in accordance with the experimental results as well
as with other model predictions. The predicted properties of the $b
\bar c$ system are
expected to be supported by the future experimental observations. \\
In Table \ref{tab:9}, we present the mean square radii of $Q\bar{Q}$
(Q $\epsilon$ b, c) systems. Our predicted values are in accordance
with few available predictions for $c \bar c$ and $b \bar b$ states
available in literature. However for the $b \bar c$ system
we do not find their sizes available in literature for comparison.\\
In Table \ref{tab:10}, the average quark velocity at the ground
state as well as at different excited states
$\left<v^{2}_{q}\right>^\frac{1}{2}$ of $Q\bar{Q}$ (Q $\epsilon$ b,
c) systems are listed for the potential index $\nu=0.7$ to 1.3. The
present results are in unit of the velocity of light. Our results
for $c \bar c$ and $b \bar b$ systems are in accordance
 with the existing values reported by others \cite{Gunar1997}up to $4S$ states.
 As expected, the quark velocity
 $\left<v^{2}_{q}\right>^\frac{1}{2}$ increases with higher excited
 states. However, it is also been observed that with increase in the
 potential index $\nu$, the quark velocity also increases (See Table
 \ref{tab:10}). It corresponds to strong binding and fast motion
 unlike the usually expected case of strong binding and slow motion.
 The predicted quark velocity of $c \bar c$ system in $6S$ states
 for the potential index 1.3 is interesting as it exceeds unity.
 Probably it may be the indication of the limit at which the $c \bar c$
 can excite. It is also supported by the fact that there exist
 little experimental evidence for the higher excited states of $c \bar
 c$ systems beyond $4S$ level. In this potential index of 1.3 the
 quark velocity approaches the velocity of light from its $4S$ state
 onwards (0.8c) warranting the relativistic approaches to study this
 states and beyond. For the choices $\nu < 1.3$, such problems do not seem to be
 important even up to the $6S$ states.\\
 In the case of $b \bar b$ systems up to $6S$ states for all the
 potential index studied here suggest the validity of
 nonrelativistic treatment. The $b$-quark/antiquark velocities up to
 $6S$ states obtained here for all the choices of the potential
 index $0.7\leq\nu\leq1.3$ lie below $0.3c$. Thus supporting the existence
of higher excited states for $b \bar b$ system compared to $c \bar
c$ system observed experimentally. In the case of $b \bar c$ system,
we have computed the velocity of $c$-quark as well as that of the
$b$-quark at different excited states. The charm quark in $b \bar c$
system seemed to move faster than its counter part in $c \bar c$
system, while the  $b$-quark in $b \bar c$ system moves slower than
that in $b \bar b$ system. Also, the importance of relativistic
effects to the motion of $c$-quark is evident for the study of its
excited states beyond $2S$ level as per the velocity predictions by
the choices of power index above $0.9$. This observation in our
present study also support the fact that the higher excited levels
will be loosely bound and may not be formed to be seen
experimentally. Over and above the predicted values of
$\left<v^{2}_{q}\right>^\frac{1}{2}$ would be useful in the study of
the decay
properties of $Q \bar Q$ systems using NRQCD formalism. \\
Our computed values of the di-gamma and leptonic decay widths with
and without the radiative corrections are shown in Tables
\ref{tab:11} and \ref{tab:12} respectively. Our predictions for $c
\bar c \rightarrow \gamma \gamma$ are in good agreement with the
experimental result for the potential index $\nu=1.1$ to $1.3$, with
out the radiative corrections. But, in the case of $b \bar
b\rightarrow \gamma \gamma$ we find our predictions with the
radiative correction are in accordance with the values reported by
others \cite{Kim2005,EbertMod2003}. In the case of leptonic decay
widths, our predictions $\Gamma_{VW}$ for both $c \bar c$ and $b
\bar b$ systems are found to be slightly over estimated in the same
range of potential index, $1.1\leq \nu \leq1.3$ and that with the
radiative corrections, $\Gamma_{ll}$ are under estimated. If may be
the indication of the fact that these decay of quarkonia occur not
at zero separation of the quark and antiquark but at some finite
separation. We must also look into the various aspects of the decay
of quarkonia discussed within the NRQCD
like formalism. We envisage such attempts for our future works.\\
We further conclude here that the present study of the properties of
$Q \bar Q$ ($Q\  \epsilon \ b, c$) systems based on the
non-relativistic coulomb plus power potential with the power index
ranging from 0.1 to 1.5 using numerical approach to solve the
Schr\"{o}dinger equation is an attempt to understand the exact
nature of the inter-quark potential and their parameters that
provided us the spectroscopic properties as well as the decay
properties of the $Q \bar Q$ system. We observe that most of the
properties of the $Q \bar Q$ systems predicted with the potential
index in the range of $0.7 \leq \nu \leq 1.3$ are in good agreement
with the existing experimental results as well as
with other theoretical model predictions.\\
\\
 \textbf{Acknowledgement:} Part of this work
is done with a financial support from DST, Government of India,
under a MajorResearch Project \textbf{SR/S2/HEP-20/2006}.
 We would like to thank Wolfgang LUCHA (Vienna) and Franz F. Sch\"{o}berl (Vienna) for providing the Mathematica code
of numerical solution of two body Schr\"{o}dinger equation. \\\\
\\\textbf{Refrences}\\

\end{document}